# Kinematics in Vector Boson Fusion

Dan Green

Fermilab

March, 2006

**Abstract:** The vector boson fusion process leads to two forward/backward jets (tag jets) and the produced state, a Higgs boson in this case, moving slowly in the p-p C.M. frame at the LHC. For the case of Higgs decaying to W+W (W*) with Higgs mass below 180 GeV, the W bosons have low momentum in the Higgs C.M. For the case of W leptonic decays, this fact allows for an approximate reconstruction of the two final state neutrinos. In turn, those solutions then provide additional kinematic cuts against background.

**Introduction:**

The vector boson fusion (VBF) production of the Higgs boson at the CERN Large Hadron Collider (LHC) is considered to be one of the best ways to search for the Higgs at low masses through the subsequent decay into W + W (W*) with both W bosons decaying leptonically [1].The process leads to two jets at small angles to the beams of protons. They are the "tag jets", which remain after an initial state quark emits a virtual W. The incident W pairs fuse to form a Higgs boson that subsequently decays into W pairs. The final state particles of the Higgs, as seen in Eq.1, are two leptons and two neutrinos.

$$u + d \rightarrow d + H + u$$
$$H \rightarrow W + W(W^*) \rightarrow \ell_1 + \nu_1 + \ell_2 + \bar{\nu}_2 \quad (1)$$

In general the final state cannot be fully reconstructed as only the two tag jets and the two leptons have their vector momenta reconstructed in the LHC detectors. However, in the special case where the Higgs mass is low, less than about 180 GeV, the W+W(W*) decay products of the Higgs have small momenta in the Higgs C.M. frame. Under the assumption that the momenta are, in fact, zero the final state neutrinos can be completely reconstructed and useful kinematic cuts can be derived on the basis of that reconstruction.

**Kinematics:**

The initial state has transverse momentum approximately equal to zero. Therefore, the neutrinos in the final state have a total transverse momentum that can be solved for using the measured momenta of the tag jets and the leptons.

Assuming small masses for the tag jets, leptons and neutrinos, there are six unknown variables – the momentum components of the two neutrinos, $\vec{P}_{\nu 1}, \vec{P}_{\nu 2}$. Two constraints come from the assumption that the initial state transverse momentum is zero, see Eq.2. If it is assumed that the process goes through a light Higgs decaying into a W plus a W (W*), then the momentum of the W in the Higgs C.M. is small. In that case the vector momenta of the two W bosons, $\vec{P}_{W1}, \vec{P}_{W2}$, are roughly equal in the lab (LHC) frame. Those five constraints allow one to solve for the individual transverse momentum components of the two neutrinos and to relate the longitudinal components of the two neutrinos.

$$[\vec{P}_{\nu 1} + \vec{P}_{\nu 2}]_T = -[\vec{P}_{tj1} + \vec{P}_{tj2} + \vec{P}_{\ell 1} + \vec{P}_{\ell 2}]_T$$
$$\vec{P}_{W1} = \vec{P}_{W2} = \vec{P}_{\ell 1} + \vec{P}_{\nu 1} = \vec{P}_{\ell 2} + \vec{P}_{\nu 2} \quad (2)$$
$$M_{\ell \nu}^2 = [(P_\ell + P_\nu)^2 - |\vec{P}_\ell + \vec{P}_\nu|^2]$$

The tag jets are labeled 1 and 2, tj1 and tj2. Number 1 is the most forward going. The leptons are labeled 1 and 2, $\vec{P}_{\ell 1}, \vec{P}_{\ell 2}$. Lepton number 1 is the most forward going. A sixth constraint arises by requiring one of the measured leptons and one of the neutrinos to have the W mass. That constraint allows one to solve, quadratically (see Eq.2), for the longitudinal momentum of one of the neutrinos. The second longitudinal momentum then follows from the assumption of equal momenta for the two W bosons.

The first lepton is tried initially. There are either two real roots of the quadratic constraint or a complex pair. If the solutions are complex then the other lepton is tried. If it too is complex then the real part is chosen to be the neutrino longitudinal momentum. If the solutions are real, then the one is chosen that has a longitudinal momentum closest to the chosen lepton longitudinal momentum.

Having solved for the neutrino longitudinal momentum by constraining one of the leptons to have a W mass in combination with one of the neutrinos, the mass of the other lepton – neutrino pair can be computed. That latter mass is not constrained. In addition, both W, constrained and unconstrained, can be paired with the forward/backward tag jets. The pairing is done choosing the W with the largest longitudinal momentum and matching it to the most forward tag jet, tj1.

**Results:**

The Monte Carlo program COMHEP [2] was used to generate VBF events and top quark pair events as one source of background. For the signal 10,000 events were generated versus 100,000 background events. Decays of the W bosons were forced to be leptonic. For the top pair backgrounds, the b quarks from the t → W + b decays were assigned to be the tag jets and the W bosons were forced to decay leptonically. Thus both processes have the same final state particle content; two jets, two leptons, and two neutrinos.

Cuts were studied which arise from the newly available kinematic information. First, the "standard" cuts, [1], were made. The rapidity difference between the tag jets was required to be greater than three and the tag jet pair mass was required to be greater than 500 GeV. The results for the cuts are shown in Table 1 for the signal, denoted qqH, and the background, given as tT.

**Table 1**: Efficiency of kinematic cuts for signal, qqH, and background process, tT

| Sequential Cut | Efficiency - qqH | Efficiency - tT |
| --- | --- | --- |
| $\Delta y > 3$ | 0.91 | 0.077 |
| $M_{tj1tj2} > 500$ GeV | 0.72 | 0.0078 |
| $M_{W uncn} < 175$ GeV | 0.55 | 0.0018 |
| $M_{tj1W1} + M_{tj2W2} > 600$ GeV | 0.49 | 0.0010 |

The distributions of the kinematic variables are shown in the following figures. In each case the distributions are those which arise when all sequential cuts prior to the variable cut in question have already been made. In Fig.1 the distribution of the rapidity difference of the tag jets is shown.

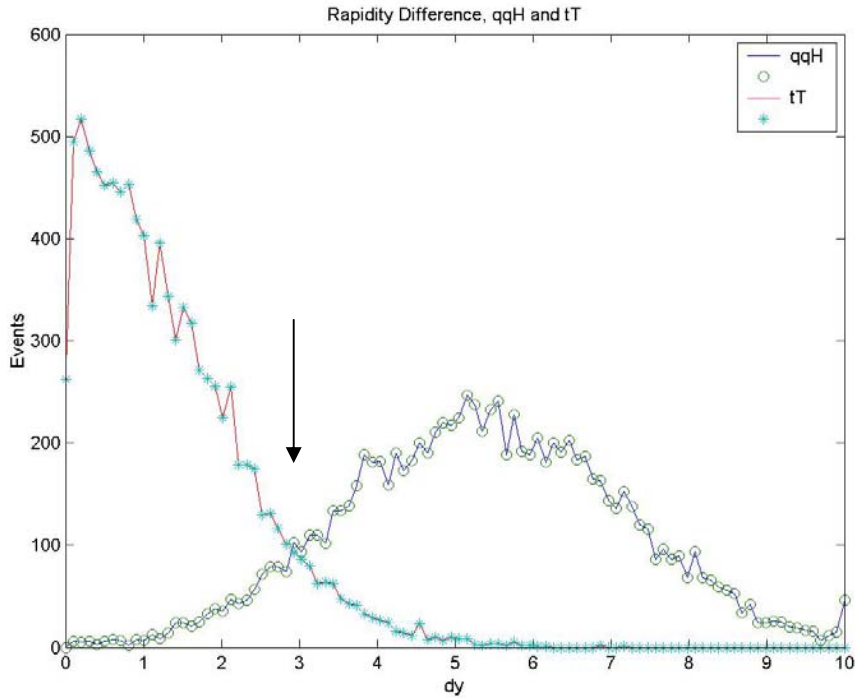

Figure 1: Distribution of the rapidity difference of the tag jets for qqH signal and tT background. The cut was subsequently set at a value of 3 as shown by the arrow.

Clearly this cut could be made rather more stringent. The mass of the tag jet pair is computed after the rapidity cut and is shown in Fig.2 for both signal and background. In all the figures the normalization of the distribution is arbitrary and only the shape is significant. A cut was set at a value of 500 GeV. These two cuts, at the generator level, are 72 % efficient for the signal and have reduced the background by a factor more than one hundred.

One of the leptons is then constrained to pair with a neutrino to give the W mass. The second lepton – neutrino pair has an unconstrained mass. To the extent that the kinematic solution is valid, one can expect that the qqH signal events, using the 180 GeV Higgs mass studied here, will display a peak near the W mass, smeared by the errors inherent in the assumptions made. The distribution of the unconstrained lepton-neutrino mass for qqH signal and tT background is shown in Fig.3. There is a prominent peak near the W mass for the signal events. Clearly, a useful cut can be made on this "unconstrained" mass (see Table 1) and a cut that the mass be less than 175 GeV was applied. This cut is about 76% efficient for the signal and reduces the top pair background by about a factor of four.

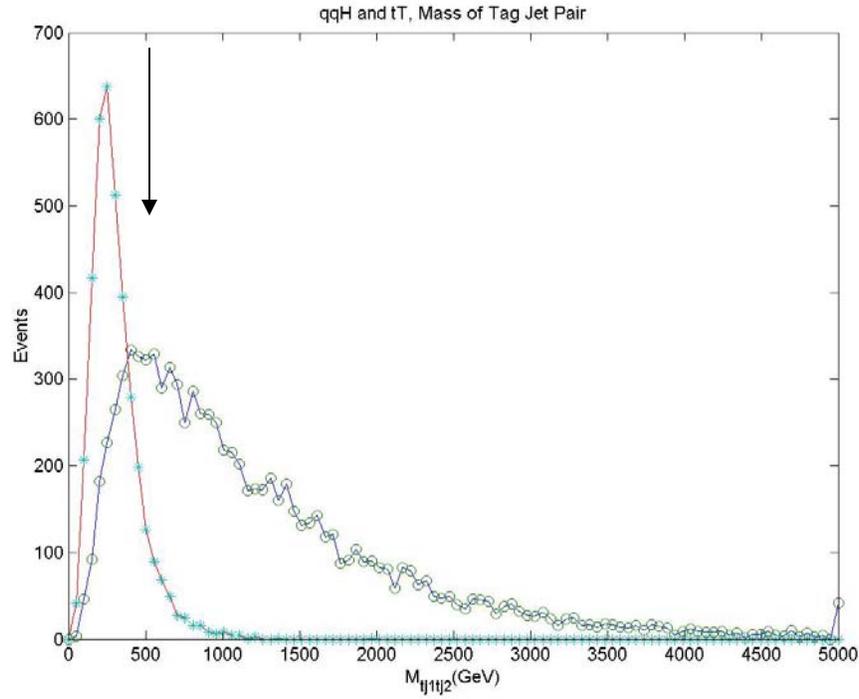

Figure 2: Distribution of the pair mass of the tag jets for qqH, o , signal and tT , * , background. The cut value of 500 GeV is indicated by the arrow.

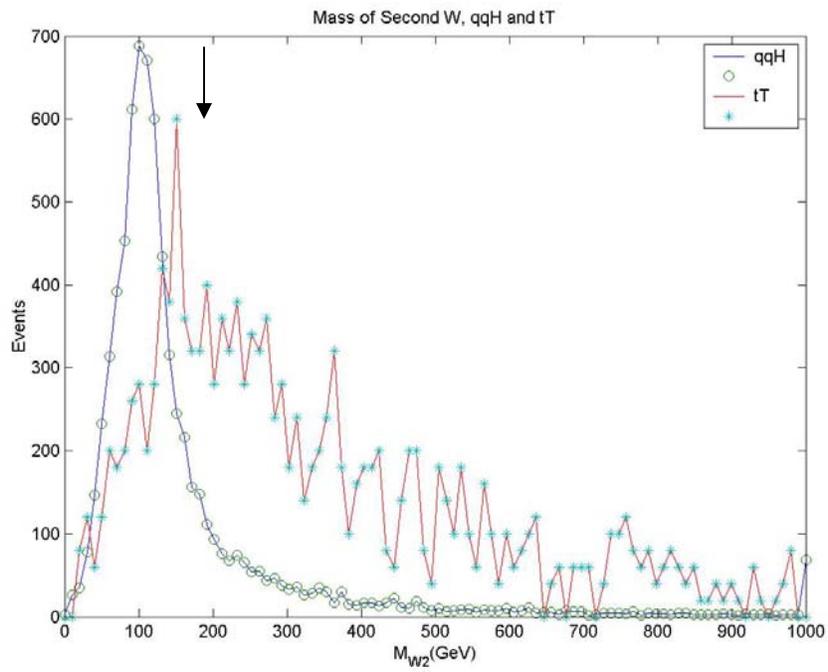

Figure 3: Distribution of the mass of the unconstrained lepton-neutrino pair for qqH signal and tT background. The cut value of 175 GeV is indicated by the arrow.

Since top pairs are a major background to VBF searches, and since the W and b are now reconstructed for the background events, a "top veto" was studied. For the top pair background the reconstructed W paired with the tag jet would, in the absence of reconstruction errors, give a top mass in both pairings. The distribution of W plus tag jet mass for both qqH signal and tT background is shown in Fig.4. Indeed, there is a low mass enhancement for the tT events which is the top mass smeared by the approximations made in the kinematic reconstruction. The qqH signal events tend to give larger masses, which makes a kinematic cut useful if not overwhelming.

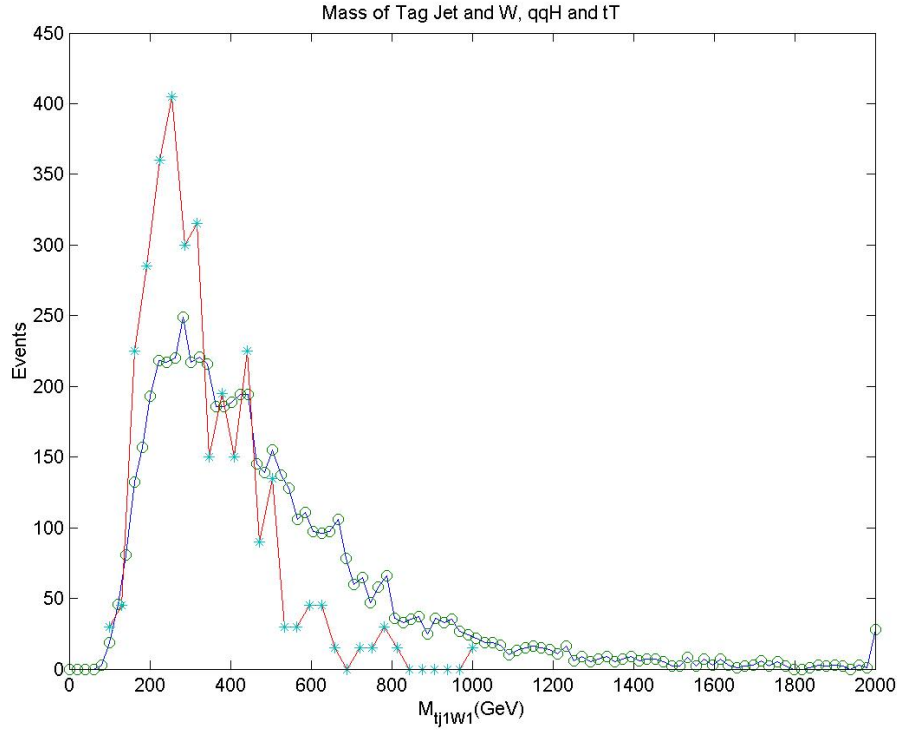

Figure 4: Distribution of the mass of the tag jet and the reconstructed W nearest to it in longitudinal momentum for qqH, o , and tT , * , events.

For the tT background the b quarks are both identified as tag jets at the generator level. Therefore, a cut on both W – tag jet masses is possible. A scatter plot of the two masses for qqH and tT events is shown in Fig.5a and Fig.5b respectively. The cut that the sum of the masses be greater than 600 GeV reduces the background by a factor (see Table 1) of 1.8 while the signal is retained at 89% efficiency.

Finally, the Higgs searches in VBF production often reduce to a "counting experiment" after all cuts have been made. In this situation the absolute rates must be well known to assess the significance of a result. Clearly, systematic effects become very important in this type of dead reckoning search. Therefore, it is useful to have a discriminating variable which possesses a differing shape for signal and background after all cuts are imposed.

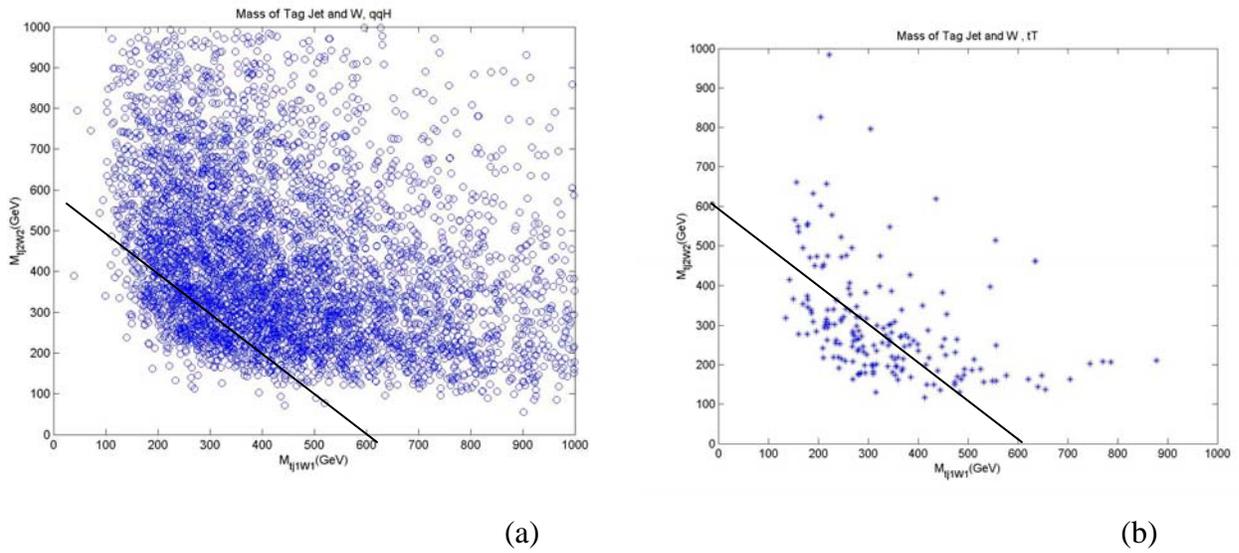

(a)                                                              (b)

Figure 5: Scatter plot of the masses of the tag jet - W pair for (a) signal, qqH, and (b) background events, tT. The lines show the cut applied on the sum of the masses.

The mass of the W pair for the generated Higgs mass of 180 GeV is shown in Fig.6. The shape of the W pair mass for the tT background is also displayed and shows a mass peaking at higher masses, near 300 GeV.

The W pair mass appears to be a useful variable in that the signal and background shapes are somewhat different after all the background removing cuts have been applied. It remains to be seen if these cuts and kinematic discriminants will be useful over a full range of Higgs masses.

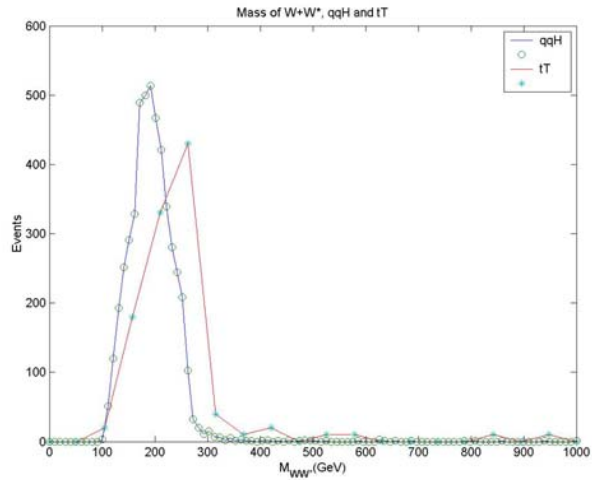

Figure 6: Mass of the W pair as reconstructed using the approximations shown in Eq.2. The qqH signal mass peaks near the generated mass of 180 GeV, while the tT background peaks at a higher mass, near 300 GeV.